\documentclass[12pt, preprint]{aastex}

\shorttitle{}

\shortauthors{Jeltema et al.}

\begin{document}

\title{ X-RAY PROPERTIES OF INTERMEDIATE-REDSHIFT GROUPS OF GALAXIES }

\author{Tesla E. Jeltema and John S. Mulchaey}

\affil{The Observatories of the Carnegie Institution of Washington}
\affil{813 Santa Barbara St., Pasadena, CA 91101}

\email{tesla@ociw.edu}

\author{Lori M. Lubin}

\affil{Department of Physics}
\affil{University of California at Davis}
\affil{One Shields Ave., Davis, CA 95616}

\author{Piero Rosati}

\affil{European Southern Observatory}
\affil{Karl-Schwarzschild-Strasse 2, D-85748 Garching, Germany}

\author{Hans B\"{o}hringer}

\affil{Max Planck Institut f\"{u}r Extraterrestrische Physik}
\affil{P.O. Box 1312, D-85741 Garching, Germany}

\begin{abstract}

We have undertaken a multiwavelength project to study the relatively
unknown properties of groups and poor clusters of galaxies at
intermediate redshifts.  In this paper, we describe the
\textit{XMM-Newton} observations of six X-ray selected groups with
$0.2<z<0.6$.  The X-ray properties of these systems are generally in
good agreement with the properties of low-redshift groups.  They
appear to follow the scaling relations between luminosity,
temperature, and velocity dispersion defined by low-redshift groups
and clusters.  The X-ray emission in four of the six groups is also
centered on a dominant early-type galaxy.  The lack of a bright
elliptical galaxy at the peak of the group X-ray emission is rare at
low-redshifts, and the other two groups may be less dynamically
evolved.  We find indications of excess entropy in these systems over
self-similar predictions out to large radii.  We also confirm the
presence of at least one X-ray luminous AGN associated with a group
member galaxy and find several other potential group AGN.

\end{abstract}


\section{ INTRODUCTION }

Most galaxies in the universe are members of groups of galaxies,
making groups an important environment for the study of galaxy
evolution (e.g. Tully 1987).  In addition to being a more common
environment for galaxies, different processes are at work in groups
versus rich clusters.  With their relatively low velocity dispersions,
groups are ideal sites for galaxy-galaxy mergers (Barnes 1985; Merritt
1985).  Many groups are also found to contain diffuse X-ray emission
(e.g. Mulchaey et al. 1993; Ponman \& Bertram 1993; Mulchaey et
al. 2003; Osmond \& Ponman 2004).  This emission is extended on scales
of hundreds of kiloparsecs, and group spectra indicate that the
emission mechanism is thermal bremsstrahlung and line emission.  The
diffuse group medium is therefore analogous to the intracluster medium
and indicates a deep potential well in these systems.  However, here
again groups may differ importantly from clusters, as
non-gravitational heating and cooling may have a larger effect in
groups.  Studies of low-redshift groups have found a steepening of the
$L_X-T_X$ relationship in the group regime (Helsdon \& Ponman 2000
a,b; Ponman et al. 1996).

The label ``group'' is used to describe a very diverse population from
loose associations of a few galaxies through poor clusters.  X-ray
emission, indicating a deep potential well, is found almost
exclusively in groups with a significant fraction of early-type
galaxies and generally in groups with a central, dominant early-type
galaxy (Mulchaey \& Zabludoff 1998; Mulchaey et al. 2003; Osmond \&
Ponman 2004).  This observation offers both clues to galaxy evolution
and a possible connection to rich clusters.  In this paper, we
concentrate on this X-ray emitting group population.

It is an open question how groups and the galaxies in them have
changed with time.  From an optically-selected group sample based on
the CNOC2 survey, Wilman et al. (2005 a,b) find that number of star
forming galaxies in groups increases with redshift to $z\sim0.5$.  On
the X-ray side, Jones et al. (2002) studied a few
intermediate-redshift groups with ROSAT and did not find evidence for
evolution in the group X-ray luminosity function.  Willis et al (2005)
find a similar lack of evolution using early data from the XMM
Large-Scale Structure Survey.  We have undertaken a program to study
in detail a sample of moderate-redshift ($0.2<z<0.6$), X-ray selected
groups.  This study includes deeper X-ray data than previous studies
of groups at these redshifts as well as HST imaging and ground based
spectroscopy.  X-ray selection provides both a method of finding
groups at higher redshifts where their sparse galaxy populations are
difficult to recognize and a well-defined selection criteria.  Our
sample and the initial spectroscopic follow-up are described elsewhere
(Mulchaey et al. 2006; hereafter Paper I).  Here we present the results of
\textit{XMM-Newton} observations of six of our groups.  Throughout the
paper, we assume a cosmology of $H_0=70h_{70}$ km s$^{-1}$ Mpc$^{-1}$,
$\Omega_m=0.27$, and $\Lambda = 0.73$.

\section{ SAMPLE }

Here we describe the follow-up with \textit{XMM-Newton} of five X-ray
selected groups of galaxies with $0.2<z<0.6$.  A sixth group,
RXJ1334+37, was found to lie off-axis in an archived \textit{XMM}
pointing.  These observations are part of a detailed X-ray and optical
study of intermediate-redshift groups; the full sample for this
project is described in Paper I.  In brief, groups were
selected from objects in the \textit{ROSAT} Deep Cluster Survey (RDCS;
Rosati et al. 1998) with luminosities in the group regime and
redshifts greater than 0.2.  These groups generally represent the more 
luminous groups in the RDCS, and the sample is fairly complete above a 
redshift of 0.3.  Figure 1 shows a comparison of the
\textit{ROSAT} and \textit{XMM} fluxes of these groups.  The fluxes
are generally similar; they are within 30\% with the exception of RXJ1648+60 where the
\textit{ROSAT} flux is most likely overestimated due to the presence
of a bright point source in the group.  Optical follow-up of these
groups includes \textit{HST} WFPC2 imaging and ground-based
spectroscopy.  Spectroscopy was obtained using the Palomar 200-inch
and Las Campanas 100-inch telescopes, and deeper spectroscopy is now
being obtained for these groups, including all of the groups with
\textit{XMM} observations, using Keck, Gemini-North and Magellan I.  The optical
data will be described in detail in future papers.  Table 1 lists
the six groups observed with \textit{XMM}, their positions, and clean
exposure times.  The positions listed are the best-fit group centers
from the surface brightness fits to the XMM data discussed in \S4.  The errors
on these positions from the spatial fitting are less than about 3'', which
is less than both the \textit{XMM} PSF and the pixel size of the PN detector (4.1'').  
We also give the group redshifts and velocity dispersions derived from
the Palomar, Las Campanas, and Keck data, as discussed in
Paper I. 

\begin{figure}
\centering
\epsscale{0.7}
\plotone{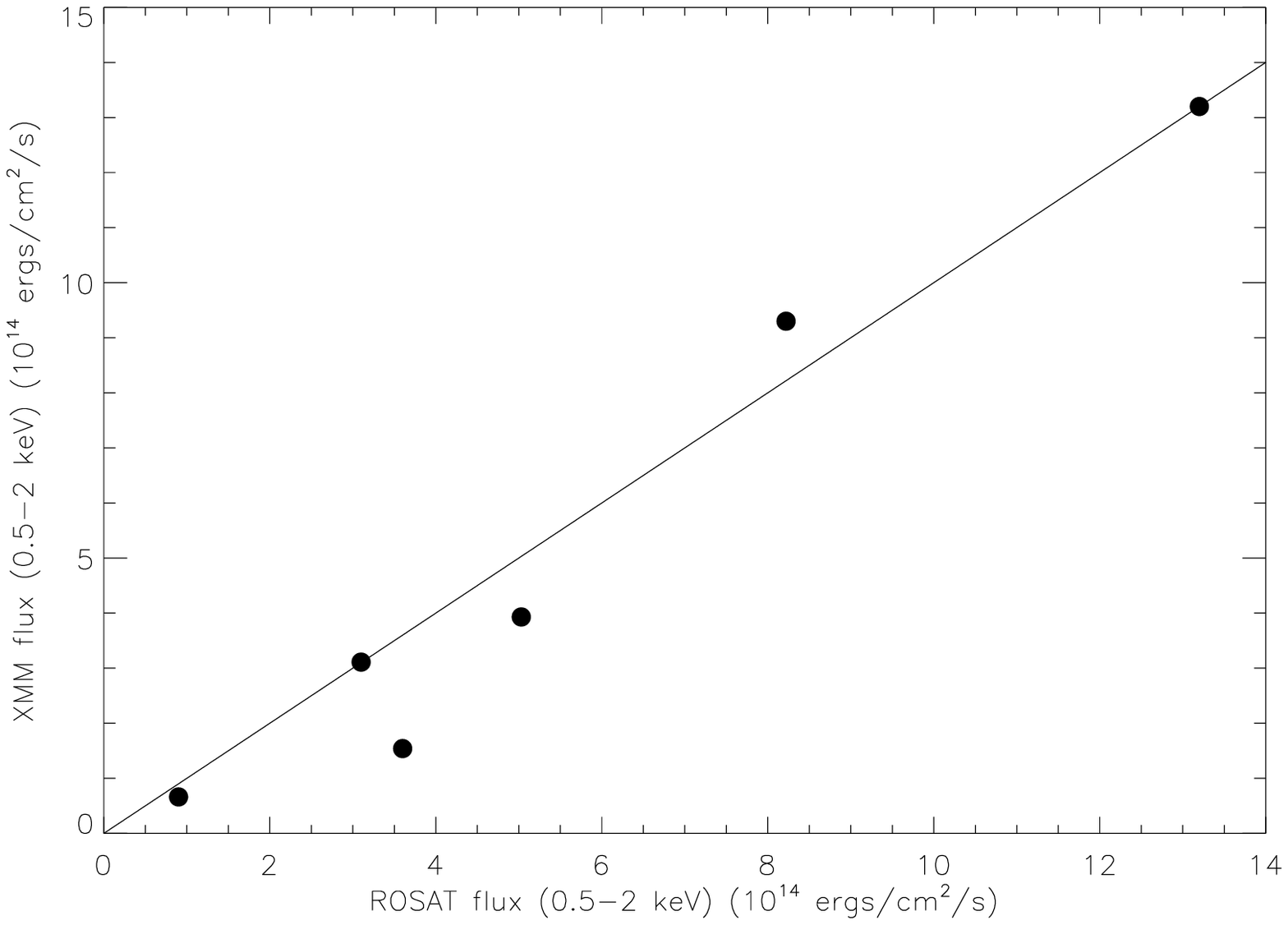}
\figcaption{ Comparison of the 0.5-2.0 keV fluxes of our groups estimated from the \textit{ROSAT} observations and determined from the \textit{XMM} observations in this paper. }
\end{figure}

\section{ DATA REDUCTION }

The six groups listed in Table 1 were observed with
\textit{XMM-Newton} between April 2001 and October 2004.  All
observations were taken in Full Frame mode with the thin optical
blocking filter.  The observation of RXJ0720+71 was split into three
observations.  We use only obs. ID 0012850701, because the other two
observations were highly contaminated with background flares.
RXJ1648+60 was observed twice with \textit{XMM}.  As discussed below,
both of these observations had significant flaring, and for this group
we merged the little time that was usable from the two observations.
RXJ1334+37 was observed approximately six arcmins off-axis in a deep
\textit{XMM} pointing available in the archive.  Similar concerns
about flaring lead us to use only the first and third of the three
observations of this pointing.

For the data reduction, we used versions 6.0 and 6.1 of the XMMSAS
software.  Observations processed with earlier versions of the
software were reprocessed using the EPIC chain tasks.  For EMOS data,
we use only patterns 0-12 and apply the \#XMMEA\_EM flag filtering,
and for EPN data, we use patterns 0-4 and flag equal to zero.  Due to
the time variability in the spectra of background flares (Nevalainen,
Markevitch, \& Lumb 2005), we filter for periods of high background in
several energy bands.  We first apply a cut on the high energy ($>$ 10
keV) count rate of 0.35 cts s$^{-1}$ for EMOS data and 1.0 cts
s$^{-1}$ for EPN data.  This cut removes the most egregious flares.
We then applied a 3$\sigma$ clipping to the source-free count rate in
three energy bands, 0.5-2 keV, 2-5 keV, and 5-8 keV.  Here time bins
(bin size of 100 secs) with rates more than 3$\sigma$ away from the
mean are removed recursively until the mean is stable.  The remaining
clean exposure times are listed in Table 1.  Flaring is significant
for RXJ1205+44 and RXJ0720+71 accounting for more than half of the
total exposure.  For RXJ1648+60, both observations are almost entirely
during periods of high background, and the filtering on high-energy
rate removes almost 60\% of the total exposure for the MOS detectors
and over 80\% of the exposure for the PN detector.  Investigation of
the 0.5-2 keV, 2-5 keV, and 5-8 keV lightcurves reveal that the
remaining exposures for both observations are also contaminated by
flares; however, very little additional time is filtered by the
3$\sigma$ clipping, because the mean rate is biased high.  In order to
make some basic measurements for this group, we proceeded with the
data remaining after the above filtering.  In our analysis, we use a
local background for both spatial and spectral fitting which should
properly account for the high background in this group.

We chose to use account for the X-ray background using a constant background 
level estimated from our observations rather than the blank-sky
event files (Lumb et al. 2002; Read \& Ponman 2003).  These
files were processed with earlier versions of XMMSAS, and changes in the
calibration mean that these files are not applicable to our data.  To estimate
the effect of possible spatial variations in the background on our fits, 
we repeated the spatial fits in Section 4 for a couple of the groups
using the blank-fields and found that this change had little effect.

\section{ SPATIAL ANALYSIS }

For each detector, we created images in the 0.5-4.5 keV band and
binned to give 2'' pixels.  Point sources were detected in a merged
image using the SAS task ewavelet and excluded from the analysis using a 25'' radius region.
Figures 2-7 show the contours from the adaptively smoothed
MOS1+MOS2+PN image of each group, before exposure correction, overlaid
on the \textit{HST} WFPC2 image.  Here point sources have been removed
before smoothing and source holes filled using the CIAO tool dmfilth.

\begin{figure}
\centering
\epsscale{0.53}
\plotone{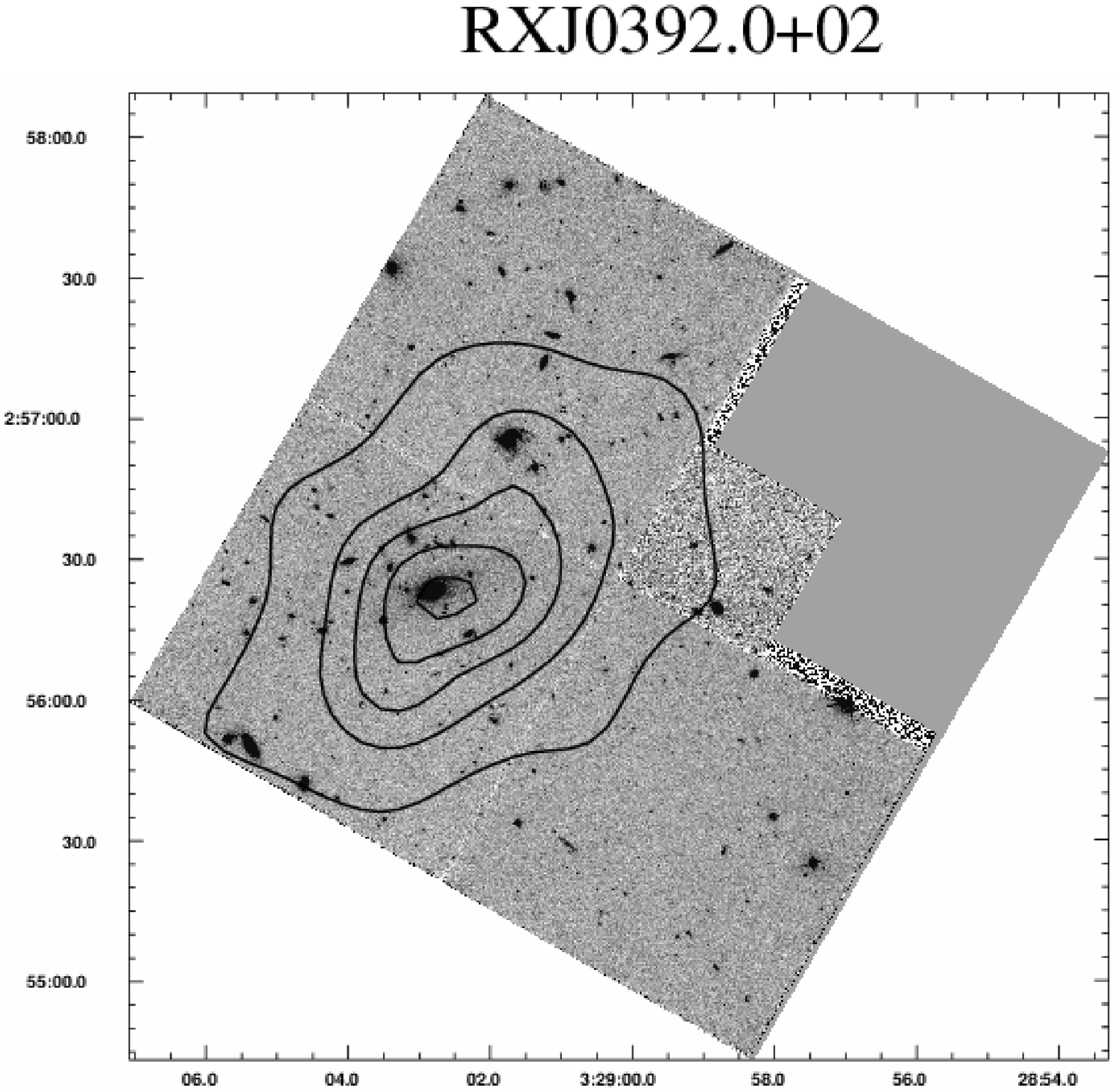}
\figcaption{ Contours from the smoothed MOS1+MOS2+PN image of RXJ0329+02 overlaid on the \textit{HST} WFPC2 image.  The X-ray contours are in the 0.5-4.5 keV band, and the image was smoothed with asmooth after the removal and filling of point source regions. The X-ray contours are linearly spaced between roughly the X-ray peak value and a bit above the background level; they were chosen to show the important group structures but not possibly spurious low surface brightness features. The WFPC2 image was taken with the F702W filter and with a total exposure of 10400 secs. }
\end{figure}

\begin{figure}
\centering
\epsscale{0.53}
\plotone{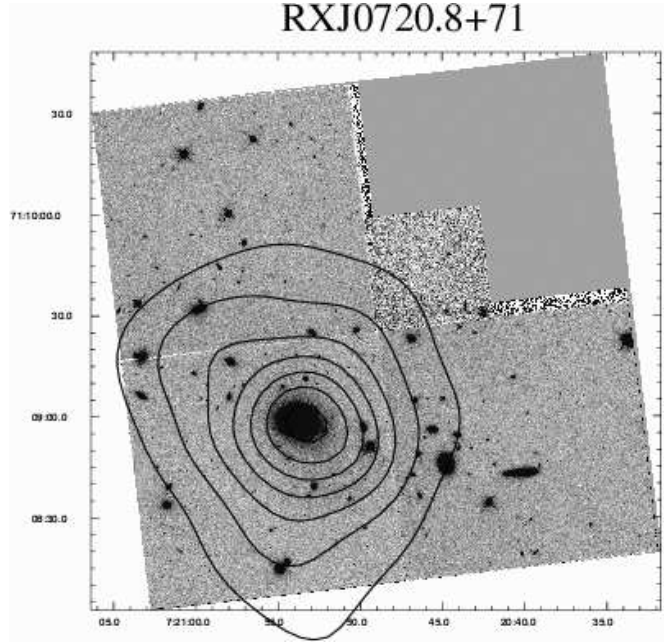}
\figcaption{ Same as Figure 2 for RXJ0720+71. The total exposure of the WFPC2 image is 5200 secs. }
\end{figure}

\begin{figure}
\centering
\epsscale{0.53}
\plotone{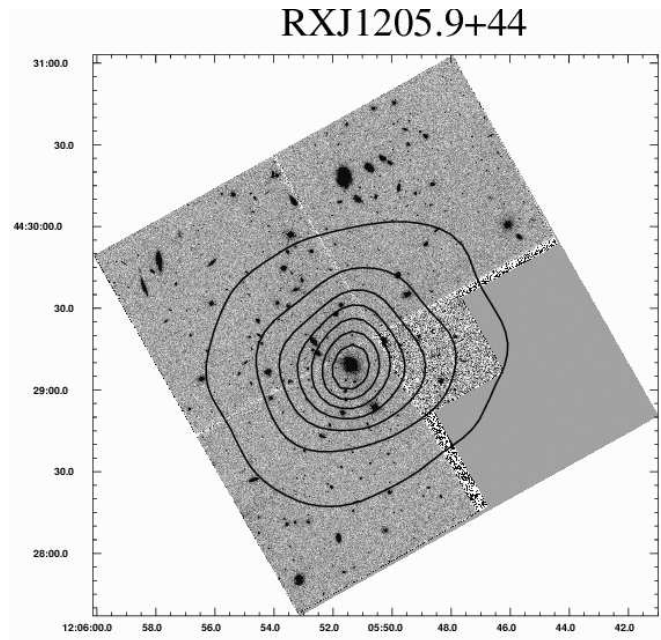}
\figcaption{ Same as Figure 2 for RXJ1205+44. The total exposure of the WFPC2 image is 7800 secs. }
\end{figure}

\begin{figure}
\centering
\epsscale{0.53}
\plotone{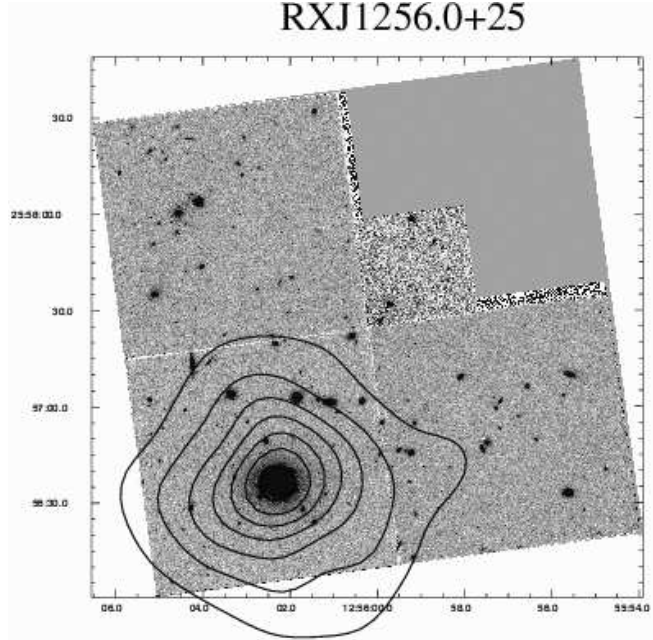}
\figcaption{Same as Figure 2 for RXJ1256+25. The total exposure of the WFPC2 image is 4400 secs. }
\end{figure}

\begin{figure}
\centering
\epsscale{0.53}
\plotone{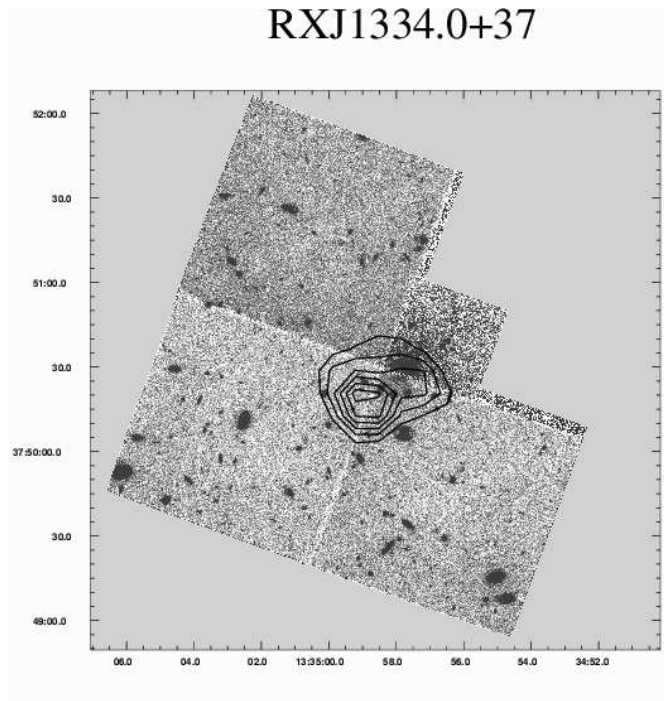}
\figcaption{ Same as Figure 2 for RXJ1334+37. A PN chip gap limits the detection of X-ray emission on the eastern side of this group. The total exposure of the WFPC2 image is 7800 secs. }
\end{figure}

\begin{figure}
\centering
\epsscale{0.53}
\plotone{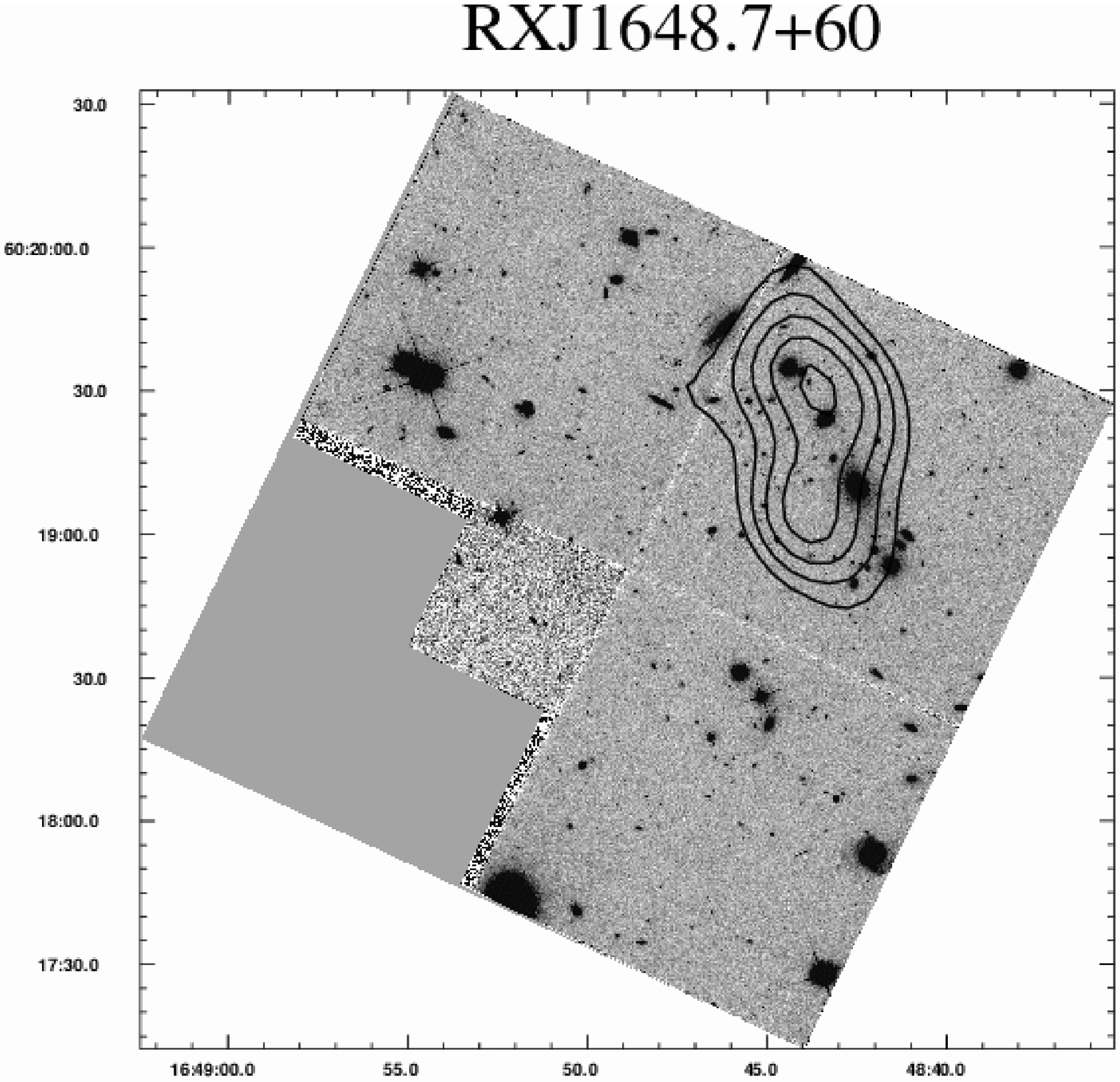}
\figcaption{ Same as Figure 2 for RXJ1648+60. This group has very few counts after flare filtering, so one should be careful when drawing conclusions from the exact shape and position of the X-ray contours.  However, the X-ray emission does align with several bright group galaxies. The total exposure of the WFPC2 image is 7800 secs. }
\end{figure}

We correct for vignetting using an exposure map, which was found to
give a similar correction to the photon weighting method (Arnaud et
al. 2001) in this energy band.  The exposure-corrected images were
then merged to create a MOS1+MOS2+PN image of each group.  This image
was normalized by the average of the exposure times to recover the
approximate number of counts.  We also created mask images to exclude
regions like chip gaps and bad pixels which had very little or no
exposure.  For RXJ1648+60 and RXJ1334+37, we combined the two
observations by first merging the exposure-corrected images,
normalized by the ratio of the exposure time to the total exposure
time, for each detector separately and then adding the MOS1,
MOS2, and PN images.

We fit the group emission to a two-dimensional $\beta$-model, of the
form
\begin{equation}
S(r) =  S_0 \left(1+ \left(\frac{r}{r_{core}}\right)^2\right)^{ -3 \beta + 0.5 },
\end{equation}
using SHERPA.  The number of counts per pixel are small, so we fit
using the maximum-likelihood based Cash statistic (Cash 1979).  The
model is convolved with an image of the \textit{XMM} point spread
function (PSF).  PSF images for each detector were created using the
EXTENDED model in calview at 2.5 keV, and in the case of RXJ1334+37,
PSF images were created for the proper off-axis distance.  For each
group, these images were weighted by the ratio of the total count rate in
each detector in the group region and merged to create a single PSF
image.  We account for the X-ray background by including a constant
background component in the model.  Groups were fit in a circular 
region centered on the group centroid with typically a radius of 
90 pixels (180''), but the region sizes varied between 30 and 120 pixels
depending on the group.  As we included the background in the fit, 
regions were chosen to be large enough to extend out to the background level.
For RXJ1648+60, the region chosen for the spatial fitting contained
approximately 280 net groups counts; all other groups contained at least
1000 net counts.

For each group, we fit the X-ray emission using both circular and
elliptical $\beta$-models, except for RXJ1648+60 which has very few
counts after flare filtering.  Even fixing the ellipticity at zero for
this group we can derive only limits on $r_{core}$ and $\beta$.  For
RXJ1334+37, the reduction in effective area at the off-axis position
of the group also leads to somewhat limited statistics, and the group
is located very close to a PN chip gap.  In order to obtain a
reasonable fit to a $\beta$-model for this group, it was necessary to
fix the position at the X-ray peak (also the centroid) and the
background at a local background estimate.  The results of the
elliptical fits and the limits for RXJ1648+60 are listed in Table 2. 
The derived core radii are significantly larger than the \textit{XMM} PSF,
so we are not limited by the \textit{XMM} spatial resolution.
RXJ0329+02 and RXJ1334+37 have significant ellipticity; the other
groups are constrained to be fairly round.  Due to the limitations of 
the statistics, we do not fit two-component models to our groups.

Studies of low-redshift groups find them to have generally lower $\beta$ values than the typical $\beta \approx 0.67$ found for rich clusters (Arnaud \& Evrard 1999; Mohr, Mathiesen, \& Evrard 1999).  Mulchaey et al. (2003) find a mean $\beta$ of $0.47\pm0.16$ for their group sample, and Osmond \& Ponman (2004) find a median value of 0.45 and a maximum value of 0.58 for the GEMS groups.  The spatial fits to our groups vary significantly.  However, three of the five groups for which we have measured $\beta$'s have $\beta$ values higher than the maximum of 0.58 found for the GEMS groups, and we find an average $\beta$ of 0.74.  Our groups also have significantly larger core radii than the GEMS sample.  Even when considering the correlation of $\beta$ and core radius, the 90\% confidence contours show that for three groups our $\beta$ values are constrained to be higher than 0.52-0.63.  There are several effects that could lead to the difference in $\beta$-model fits.  Our groups are generally hotter than the groups in the two low-redshift samples mentioned, and we detect our groups to a generally larger fraction of the virial radius than was used for the cooler low-redshift groups.  Willis et al. (2005) also find relatively low $\beta$'s for moderate-redshift ($0.29<z<0.44$) groups detected in the \textit{XMM} Large-Scale Structure (LSS) survey.  For the six groups with measured temperatures and temperatures below 3 keV, they find an average $\beta$ of 0.53 and a maximum of 0.67.  However, fewer than 200 counts are detected from some of these groups, and so we are again probing a significantly larger region in our groups.

\section{ SPECTRAL ANALYSIS }

For spectral fitting, we defined the extent of each group to be the
radius at which the surface brightness reaches 20\% of the background.
This radius, $r_{ext}$, was determined from the one-dimensional
surface brightness profile and using the best-fit background level
from the 2D $\beta$-model.  We extract spectra separately for each
detector, and we extract local background spectra from an annular
region of similar area outside where the surface brightness reaches
the background level.  As mentioned above, due to changes in the
calibration the blank-field backgrounds are not applicable to our
data.  Point sources were again excluded from the analysis.  The
spectra are grouped to give a minimum of 25 counts per bin, and RMFs
and ARFs are created for each detector using rmfgen and arfgen.  Our
spectral extraction regions are not much larger than the scale of
instrumental variations included in the ARFs ($\sim$ 1 arcmin), and we
found a flat detector map to be sufficient for ARF creation.

For each group, the spectra from the three detectors were jointly fit
to an absorbed mekal model in the 0.5-5 keV band.  We ignore energies
between 1.45 and 1.55 keV to exclude the Al K instrumental lines.  For
RXJ1648+60 and RXJ1334+37, the six spectra resulting from the two
observations are all jointly fit, and for RXJ1648+60, we use a broader
energy band of 0.5-8 keV to improve the signal.  We fix the absorbing
column at the galactic value (Dickey \& Lockman 1990) but allow both
the temperature and metallicity to vary.  Metallicities are relative 
to the abundances of Anders \& Grevessa (1989).  In the case of RXJ1648+60,
the metallicity was fixed at 0.3 solar.  The results of these fits are
shown in Table 3 along with the 90\% confidence limits.  Also listed
are the unabsorbed, bolometric luminosities (0.01-100 keV) determined from the
spectral fits.  The luminosity errors include both the uncertainty in
the temperature and the metallicity.  For RXJ1648+60, we varied the
metallicity between 0.0 and 1.0 solar to bound the luminosity error,
and an upper limit of 1.0 solar was also used for RXJ1334+37.  We
correct our luminosities for the flux lost due to point source removal
using the best-fit $\beta$-model.  In Sherpa, we create a model image
of each group and calculate the count rate in the spectral extraction
region both with and without the point source regions.  The
luminosities are then corrected by the ratio of these two count rates.

Our groups have temperatures around 2 keV placing them in the massive
group or poor cluster regime of galaxy associations.  The best-fit
temperature for RXJ1648+60 is lower at around 1 keV, but the large
errors associated with the small number of group counts mean that it
is consistent with the other groups.  The best-fit metallicities of
our groups span a large range in metallicity and generally have large errors.  
One of the groups, RXJ1256+25, is constrained within the 90\% confidence limits 
to have a fairly low metallicity, below 0.25 solar.  Only
RXJ1334+37 and RXJ1205+44 have metallicities greater than zero within
the 90\% confidence limits, and only RXJ1205+25 has a significant
metallicity with a lower limit of 0.27.  Deeper observations are
needed to truly constrain the abundance of metals in
intermediate-redshift groups.

The spectral extraction radius varies from group to group depending on
both flux and exposure time.  In order to compare the groups at a
standard radius, we extrapolated their luminosities to $r_{500}$, the
radius at which the density is five hundred times the critical
density.  We estimate $r_{500}$ for each group in the assumed
$\Lambda$CDM cosmology using the best-fit values of $\beta$, core
radius, and temperature.  Assuming isothermality and a $\beta$-model
surface brightness distribution, the total mass within a radius $r$ is
given by
\begin{equation}
M_{tot}(<r) =  \frac{3 \beta T r_{core}}{G \mu m_p}\frac{x^3}{1+x^2},
\end{equation}
where $x = r/r_{core}$, $\mu=0.6$ is the mean molecular weight, $G$ is the gravitational constant, and $m_p$ is the proton mass.  The ratios of our spectral extraction radii to $r_{500}$ range between 0.45 and 1.05, so we are probing a reasonable fraction of these groups.  Luminosities were then extrapolated using the best-fit $\beta$-model.  For RXJ1648+60, for which we were unable to obtain an accurate fit to a $\beta$-model, we use the average parameters from the other five groups, $r_{core} = 160$ kpc and $\beta = 0.74$.  These luminosities are listed in Table 3 along with the values of $r_{500}$; errors in luminosity were determined from the spectral errors alone and do not include the uncertainty in the $\beta$-model parameters.  Most of the group luminosities increase only slightly when extrapolated to $r_{500}$, and the correction is always less than a factor of 1.7.

\section{ COMPARISON TO LOW-REDSHIFT GROUPS }

\subsection{ Brightest Group Galaxies }

Investigation of Figures 2-7 reveal that with the exception of
RXJ1648+60 and RXJ1334+37
the groups are centered on dominant early-type galaxies.  In
particular, the two lowest-redshift groups, RXJ0720+71 and RXJ1256+25,
have large central galaxies.  In both of these groups, the central
object is composed of three components, and in RXJ0720+71 the two
brightest components are consistent with having the same radial
velocity (Paper I).  The multiple nuclei in these systems are only
separated by a few arcsecs, so the \textit{XMM} peak can not be 
reliably identified with a single component.  In RXJ1205+25, the brightest group
galaxy (BGG) likewise has two components (Paper I).  In
RXJ0329+02, the position angle of the central galaxy and the X-ray
emission appear to align.  This alignment can be seen in Figure 8
which shows the central contours of the best-fit $\beta$-model to the
X-ray emission overlaid on the HST image of the BGG.  From Sherpa, the
position angle of the semi-major axis of the X-ray emission is
$-59\pm6^{\circ}$, measured counterclockwise from north.  Fitting the
central galaxy with the task ELLIPSE in IRAF, we find a position angle
between $-66^{\circ}$ and $-78^{\circ}$ for all but the inner most
ellipse.  The X-ray emission therefore aligns with the BGG to within
$20^{\circ}$.  Although the position of the X-ray center from the
X-ray contours appears to be offset from the galaxy center, this
offset, which is a couple of arcsecs, is not significant.

\begin{figure}
\centering
\plotone{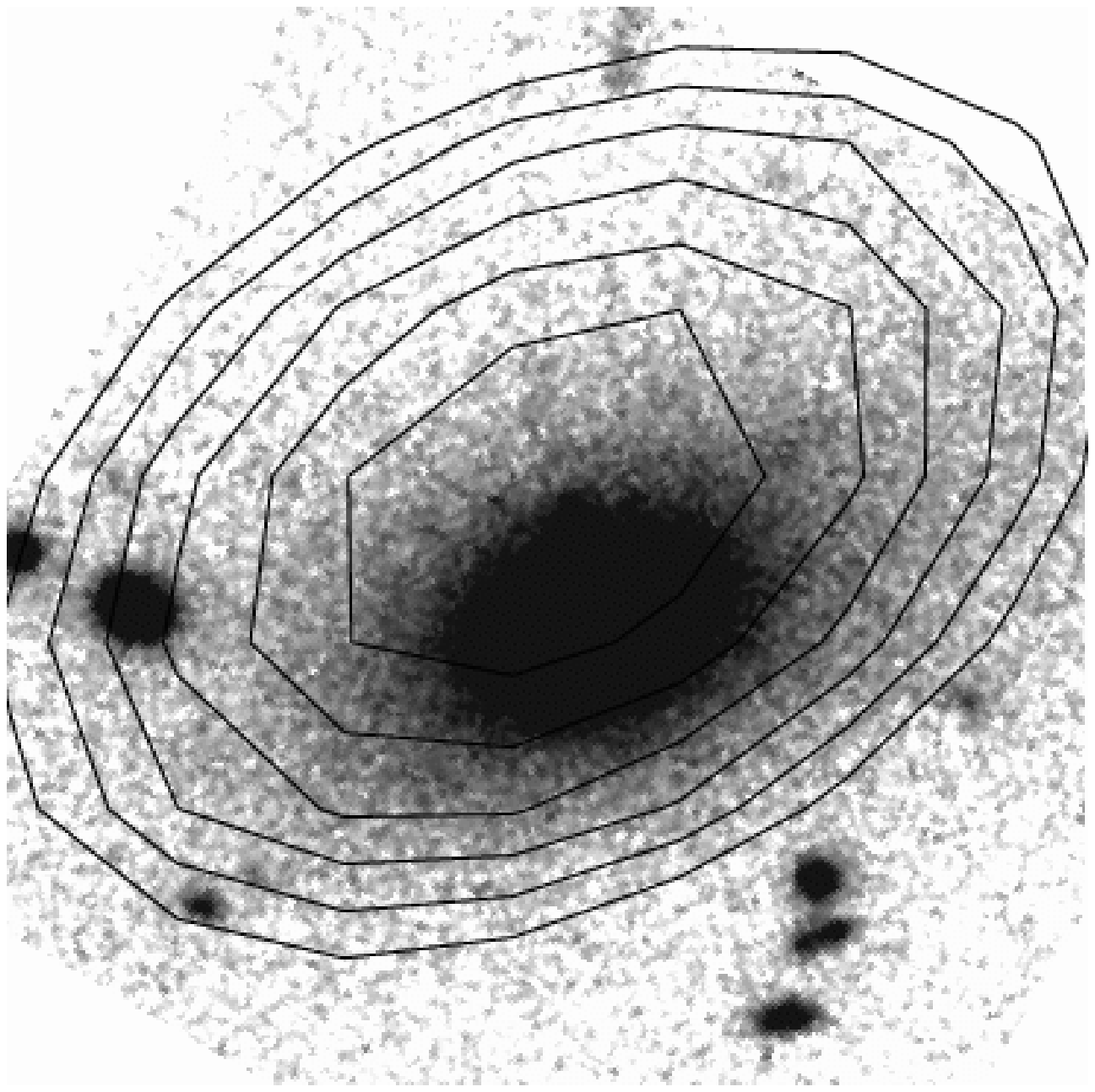}
\figcaption{ HST image of the central group galaxy in RXJ0329.0+02 overlaid with the central contours of the best-fit 2D $\beta$-model to the X-ray emission. Both the X-ray emission and the central galaxy have a similar position angle.  Although the X-ray contours appear to be offset from the galaxy center, this offset is not significant given the uncertainty in the X-ray positions. }
\end{figure}

These observations correspond well to the observed properties of
low-redshift groups.  Similar to clusters, low-redshift groups in
which X-ray emission is detected are almost all centered on dominant
early-type galaxies (Osmond \& Ponman 2004; Mulchaey \& Zabludoff
1998).  However, two of our groups do not show this feature.
RXJ1334+37 contains a dominant early-type galaxy, but the X-ray
emission is centered to the south-east of this galaxy.  RXJ1648+60
contains a string of bright group galaxies which are traced by the
X-ray emission, but from the current data the X-ray emission is not
peaked on any one of these.  This group also does not contain a
clearly dominant galaxy, but rather several galaxies with similar
magnitudes (Paper I).  Another group studied in Mulchaey et al. (2006) 
for which we do not yet have \textit{XMM} data, RXJ0210-39, also has a 
chain-like morphology and no dominant early-type galaxy.  
These groups may not be as dynamically evolved as low-redshift
groups or the other groups in our sample.  In addition, three of the
four groups in our sample with central galaxies have BGGs with multiple cores.
In contrast, in an X-ray selected sample of 19 groups at $z\le0.05$ with 
similar luminosities, described in Paper I, the X-ray emission in all
of the groups is centered on a bright early-type galaxy, and none of the BGGs
have multiple nuclei.  Therefore, X-ray luminous groups at intermediate redshifts
appear to be in an earlier stage of formation than low-redshift groups,
at least as far as their galaxies are concerned (Paper I).  The detailed morphological
content of our moderate-redshift groups will be investigated in a
future paper.

\subsection{ Scaling Relations }

It is well known that there is a strong correlation between X-ray
luminosity and temperature in clusters of galaxies, and that the
$L_X-T_X$ relation for clusters does not follow the expected $L_X
\propto T_X^2$ for self-similar systems radiating through thermal
bremsstrahlung.  This relation is instead observed to be roughly $L_X
\propto T_X^3$ (e.g. White, Jones, \& Forman 1997; Arnaud \& Evrard
1999).  The scaling between luminosity and temperature in the group
regime is less well established.  Studies of low-redshift groups
have found that the $L_X-T_X$ relation steepens significantly in the
group regime (Helsdon \& Ponman 2000 a,b; Ponman et al. 1996).

In Figure 9, we show the relationship between luminosity and
temperature for our groups compared to the low-redshift groups ($z < 0.03$) in the
GEMS sample (Osmond \& Ponman 2004).  For comparison, we also plot 
a sample of low-redshift ($z < 0.09$) clusters with ASCA temperatures and ROSAT
luminosities from Markevitch (1998).
Our groups lie in the region of this plot in between what are
typically labeled clusters and what are typically labeled groups, and 
they show good agreement with the low-redshift samples.

Figure 10 also compares the $L_X-T_X$ relation for our groups versus
the GEMS sample but with the luminosities projected to $r_{500}$.  As
with our groups, the luminosities of the GEMS groups were extrapolated
to $r_{500}$ using the best-fit $\beta$-model, and
for those groups where they were unable to fit a $\beta$-model, the
average core radius and $\beta$ were used (Osmond \& Ponman 2004).
While this extrapolation involves a certain amount of uncertainty, it
is important because the group luminosities are measured within very
different radii, and the correction can be as high as a factor of
three for the low-luminosity GEMS groups.  Also plotted are the six
groups from the XMM-LSS survey with measured temperatures below 3 keV.
These groups have redshifts between 0.29 and 0.44.  Here the solid
line shows the best-fit to the GEMS $L_X(r_{500})-T_X$, while the
dotted and dashed lines show the fits to the Markevitch (1998) cluster
sample and the GEMS plus Markevitch samples, respectively (Willis et al. 2005; 
Helsdon \& Ponman, in preparation).  As noted in 
Willis et al. (2005), the luminosities in Markevitch (1998) are quoted
within $1 h_{100}^{-1}$ Mpc not $r_{500}$, but the correction factors
are typically smaller than the 5\% calibration uncertainty in the luminosities.
In this plot,
our errors are more conservative 90\% confidence limits, and we
include the uncertainties in temperature and metallicity in the
luminosity errors, while 1$\sigma$ limits are shown for the other two
samples.  Again, our moderate-redshift groups agree with the
low-redshift $L_X(r_{500})-T_X$ relation within the errors as well 
as with the cluster or cluster plus group
relations.  The observations of the XMM-LSS survey groups are
generally not as deep as ours, but they cover a similar redshift
range.  They also appear to be consistent with the luminosities and
temperatures of our groups.  In general, the intermediate-redshift groups
scatter about both the low-redshift cluster and group $L_X(r_{500})-T_X$ fits
with possibly slightly better agreement to the clusters.  Here we have not applied an evolutionary
correction to our luminosities (Ettori et al. 2004), but in our redshift range
this correction is small.  Using $E_z^{-1}L_X \propto T_X$ would reduce our
luminosities by at most a factor of 0.75. 

\begin{figure}
\centering \epsscale{1.0} \plotone{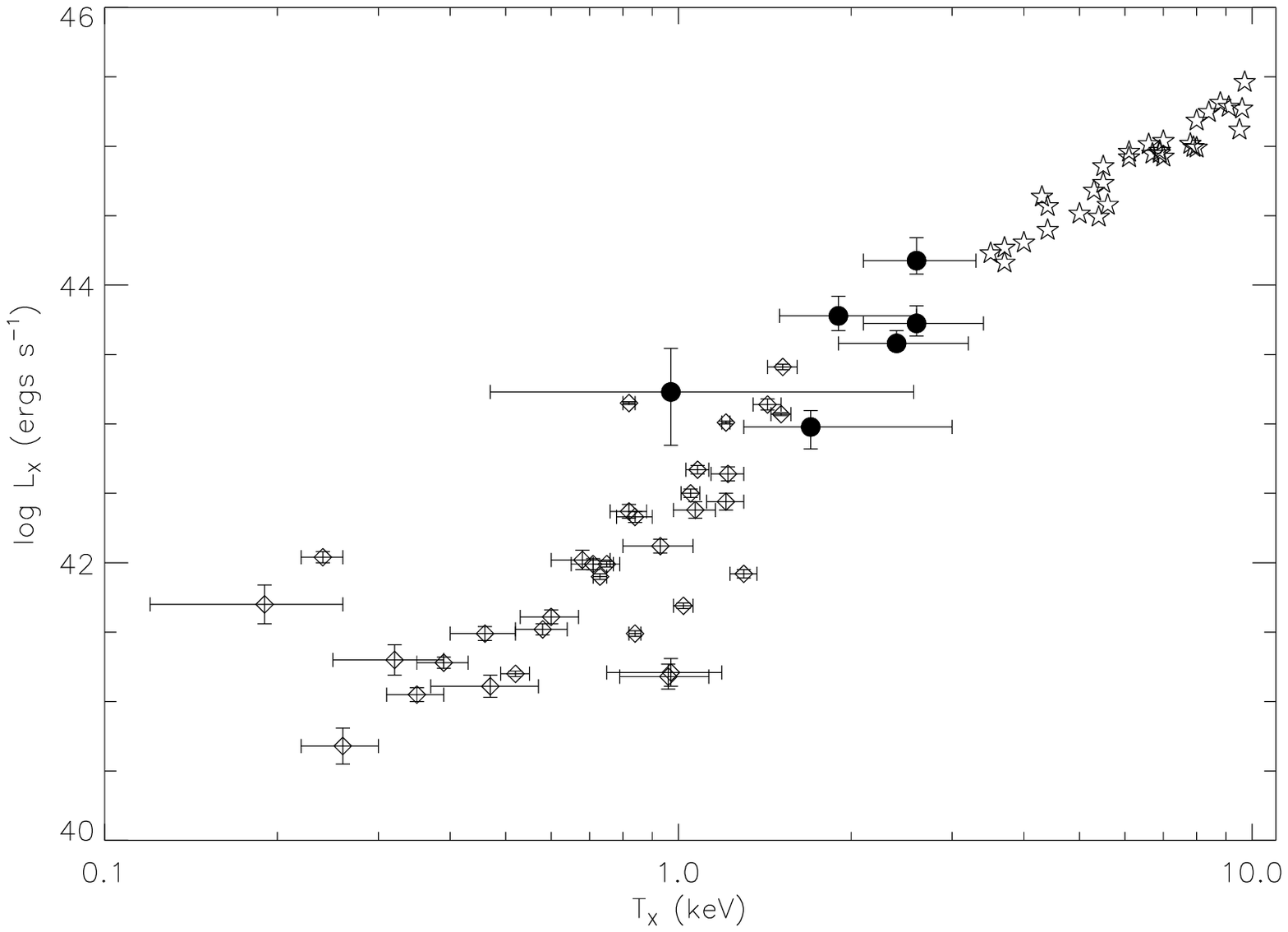} 
\figcaption{ $L_X-T_X$
relation for our groups (filled circles), the low-redshift GEMS groups
(Osmond \& Ponman 2004)(open diamonds), and the cluster sample of
Markevitch (1998) (open stars).  Luminosities are unabsorbed, bolometric
luminosities.  The error bars on our points show the 90\% uncertainty
in temperature.  The luminosity errors include both the uncertainty in
temperature and metallicity.  Errors on the GEMS points are 1$\sigma$
with the luminosity errors calculated from Poisson statistics. }
\end{figure}

\begin{figure}
\centering
\epsscale{1.0}
\plotone{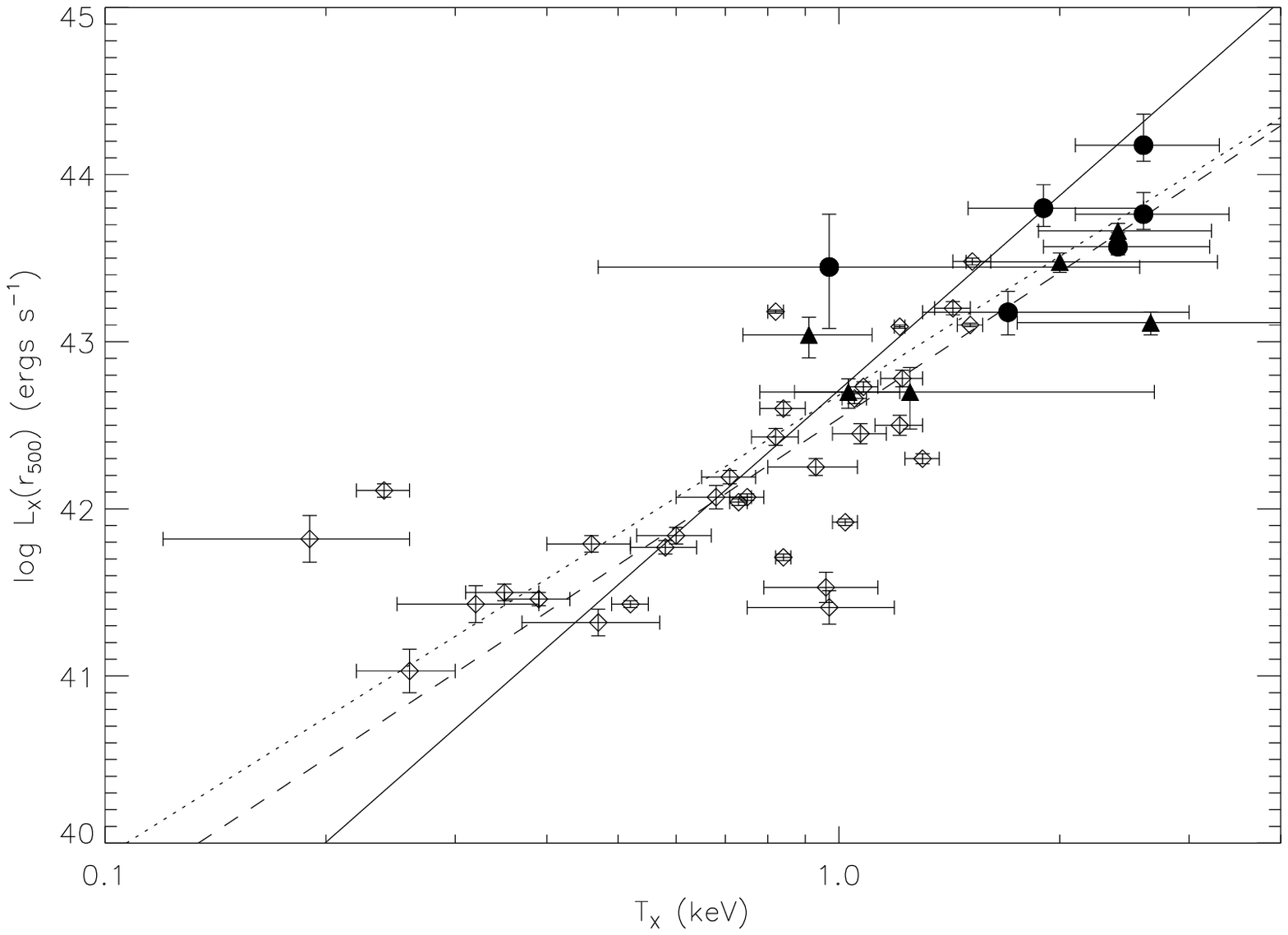}
\figcaption{ Relationship between $L_X(r_{500})$ and $T_X$ for our groups (filled circles), the low-redshift GEMS groups (Osmond \& Ponman 2004)(open diamonds), and the moderate-redshift XMM-LSS groups (Willis et al. 2005) with measured temperatures (filled triangles).  The error bars on our points show the 90\% uncertainty in temperature, and the error in luminosity from both the uncertainty in temperature and metallicity.  Error bars for the other samples are 1$\sigma$.  Also shown are fits to the GEMS groups (solid line), the Markevitch clusters (dotted line), and the combination of the Markevitch (1998) clusters and the GEMS groups (dashed line)(Helsdon \& Ponman, in preparation). }
\end{figure}

We also investigate the relationship between the velocity dispersion
of the group member galaxies and the X-ray temperature.  Figure 11
shows a comparison of our groups to low-redshift groups and clusters.
Here our groups and the XMM-LSS groups lie closer to the cluster
$\sigma_v-T_X$ relation
then they do to the fit to the low-redshift GEMS groups.
In Figure 11, it can be seen that there is a large scatter in the velocity 
dispersions of our groups.  Some of these velocity dispersions were determined 
from relatively few galaxies (Paper I).  In addition, our groups were 
selected based on X-ray luminosity versus
low-redshift group samples which are primarily optically-selected.
This X-ray selection could lead to a larger scatter in optical
properties.  In general, our groups show agreement with the cluster 
$\sigma_v-T_X$ relation, but a few fall off this relation within the errors.  In 
particular the two most discrepant points, RXJ1648+60 and RXJ1334+37,
have velocity dispersions which appear significantly low for their temperatures.
These groups also fall significantly
off the $L_X-\sigma_v$ relation (Paper I).  The spectrum
of RXJ1648+60 only has a S/N of 3.6, and the associated error
in the temperature could make it consistent with the $\sigma_v-T_X$
relation but not the the $L_X-\sigma_v$ relation.  The velocity dispersions 
for these groups were also
computed from relatively few galaxies, six and eight respectively
(Paper I).  The measured velocity dispersions for both
groups are near the theoretical lower bound of $100-200$ km s$^{-1}$
for a collapsed system (Mamon 1994), yet both show extended X-ray
emission.  Similar low velocity dispersion, over-luminous groups were
found in the GEMS sample, but at lower X-ray luminosities and
temperatures, and at least two of these are confirmed to have
significant X-ray emission from deeper Chandra observations (Helsdon,
Ponman, \& Mulchaey 2005).  Helsdon et al. (2005) propose several
possible physical effects which could lead to low velocity dispersions
including dynamical friction, tidal heating, and orientation effects.
On going deeper spectroscopy of our sample will help to determine the 
nature of these groups.

\begin{figure}
\centering
\epsscale{1.0}
\plotone{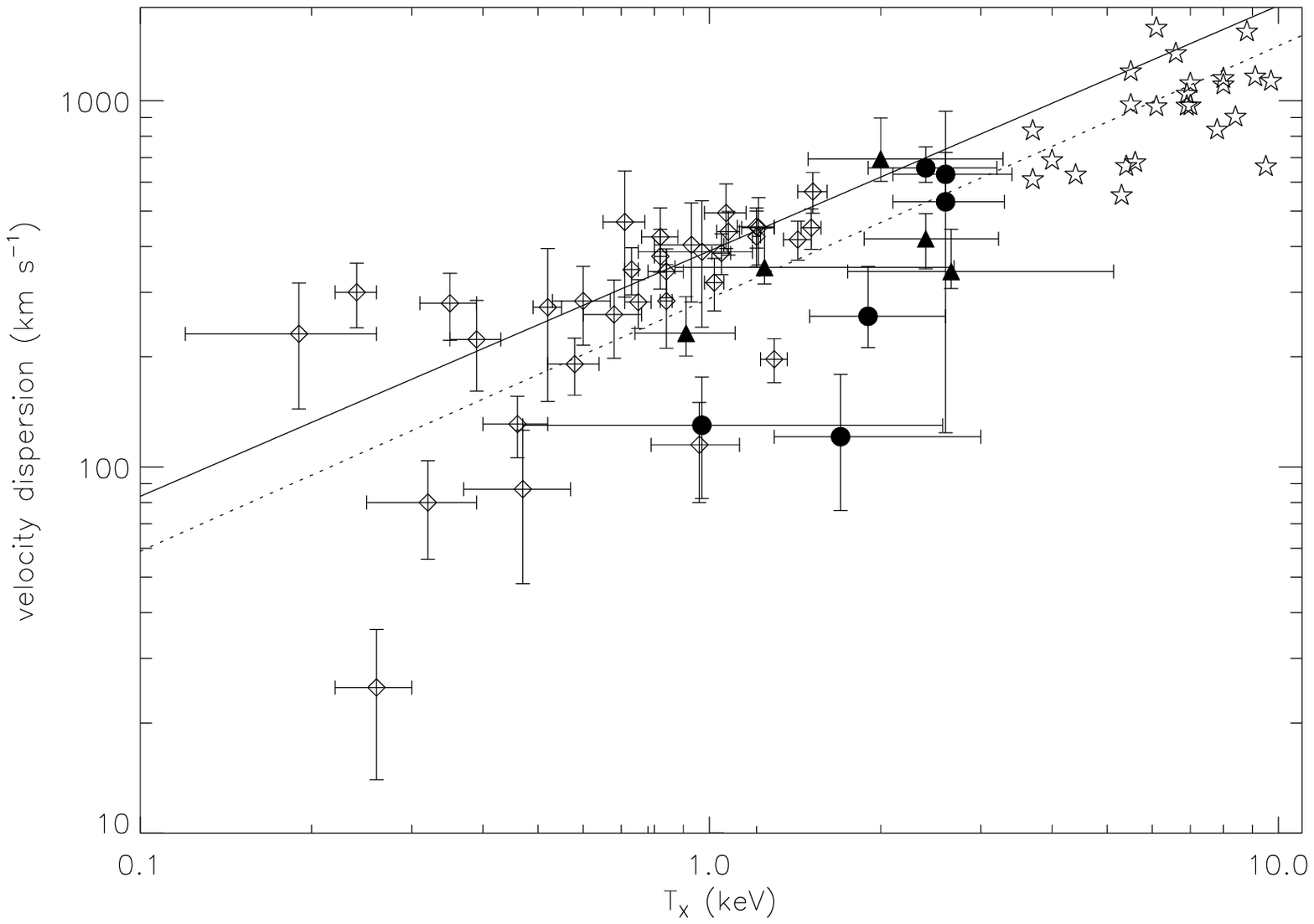}
\figcaption{ $\sigma_v-T_X$ relation for our groups (filled circles), the low-redshift GEMS groups (Osmond \& Ponman 2004)(open diamonds), the cluster sample of Markevitch (1998) (open stars), and the moderate-redshift XMM-LSS groups (filled triangles)(Willis et al. 2005).  Velocity dispersions for the Markevitch sample are taken from Horner (2001).  The error bars on our points show the 90\% uncertainty in temperature and the 1$\sigma$ errors in $\sigma_v$.  Error bars for the other samples are 1$\sigma$.  Also shown is the fit to the GEMS groups (solid line) and a fit to the Markevitch clusters (dotted line)(Helsdon \& Ponman, in preparation). }
\end{figure}

\subsection{ Entropy }

As with the $L_X-T_X$ relation, the entropy of the IGM in low-redshift
groups is observed to deviate from the self-similar expectation of a
simple linear scaling of entropy with temperature (Ponman, Cannon \&
Navarro 1999; Lloyd-Davies, Ponman, \& Cannon 2000; Finoguenov et
al. 2002; Ponman, Sanderson, \& Finoguenov 2003).  Explanations for
these deviations from self-similarity include preheating of the gas
before clusters were assembled (Evrard \& Henry 1991; Kaiser 1991;
Cavaliere, Menci \& Tozzi 1997; Balogh, Babul \& Patton 1999; Valageas
\& Silk 1999; Tozzi \& Norman 2001; Babul et al. 2002; Dos Santos \&
Dor\'{e} 2002; Nath \& Roychowdhury 2002), heating by supernova and/or
AGN (Bower 1997; Loewenstein 2000; Voit \& Bryan 2001; Nath \&
Roychowdhury 2002; Roychowdhury, Ruszkowski, \& Nath 2005), or removal
of low-entropy gas through cooling (Knight \& Ponman 1997; Bryan 2000;
Pearce et al. 2000; Muanwong et al. 2001; Wu \& Xue 2002; Dav\'{e},
Katz \& Weinberg 2002).  Using a sample of 66 systems ranging in mass
from galaxies to massive clusters, Ponman et al. (2003) find a trend
of entropy, measured at $0.1r_{200}$, versus temperature that is
shallower than the self-similar expectation of $S \propto T$.  Their
results suggest the presence of extra entropy in systems at all
temperatures relative to the hottest clusters.  Similar to Finoguenov
et al. (2002), they also detect excess entropy at a much larger
radius, $r_{500}$.  The existence of excess entropy at at large radii
conflicts with many preheating models which predict that the entropy
increase should be restricted to the central regions of groups and
clusters.

We calculate the entropies of our groups using the standard definition
of entropy as $S=T/n_e^{2/3}$.  Using the $\beta$-model
parameterization of the surface brightness, the gas density profile is
given by
\begin{equation}
n_{gas}(r) = n_{0,gas}(1+r^2/r_{core}^2)^{-3\beta/2}.
\end{equation}
The central electron density is derived from a combination of the surface brightness fit and the normalization of the spectral model.
\begin{equation}
n^2_{0,e} = \frac{1.17 D^2_A (1+z)^2 K \times 10^{14}}{EI}
\end{equation}
where $K = 10^{-14} \times \int n_p n_e dV/[4 \pi D^2_A(1+z)^2]$ is the normalization of the mekal spectrum in XSPEC, $EI= \int^{r_{ext}}_0 (1+r^2/r_{core}^2)^{-3\beta}r^2dr + \int^{10Mpc}_{r_{ext}} (1+r^2/r_{core}^2)^{-3\beta}r^2(1-$cos$\theta)dr$ with $\theta = $arcsin$(r_{ext}/r)$, $D_A$ is the angular diameter distance, and we assume $n_e=1.17n_p$ (Ettori et al. 2004).

Following Ponman et al. (2003), we calculate group entropies at both
relatively small radii ($0.1r_{200}$) and large radii ($r_{500}$).
Figures 12 and 13 show these entropies versus temperature compared to
the low-redshift groups and clusters studied by Ponman et al. (2003)
and to the self-similar prediction normalized to hot clusters.  Here
we plot $E_z^{4/3}S$ for our groups, where
$E_z=H_z/H_0=[\Omega_m(1+z)^3+\Lambda]^{1/2}$.  This scaling accounts
for the variation of the mean density within a given overdensity
radius with redshift.  We find a similar entropy excess compared to
the self-similar expectation at both radii and reasonable agreement
with the low-redshift points.  Here we have assumed isothermality, but at 
these radii we do not expect that the average emission weighted temperatures
will vary significantly from our measured temperatures, measured at between half 
of $r_{500}$ and $r_{500}$ (Rasmussen \& Ponman 2004; Ponman et al. 2003).  Rasmussen 
\& Ponman (2004) compare the entropy profiles for two groups assuming both isothermal
and polytropic gas distributions; these profiles vary by less than about 25\%.  In 
Figures 12 and 13, we plot the
1$\sigma$ errors in entropy propagated from the errors in temperature
and spectral normalization.  Additional uncertainty is present from
the $\beta$-model fits to the surface brightness and in the
calculation of overdensity radii, so these results should be
interpreted with care.  However, the entropies of our groups are consistent with
the excess entropy observed in low-redshift groups.

\begin{figure}
\centering
\epsscale{1.0}
\plotone{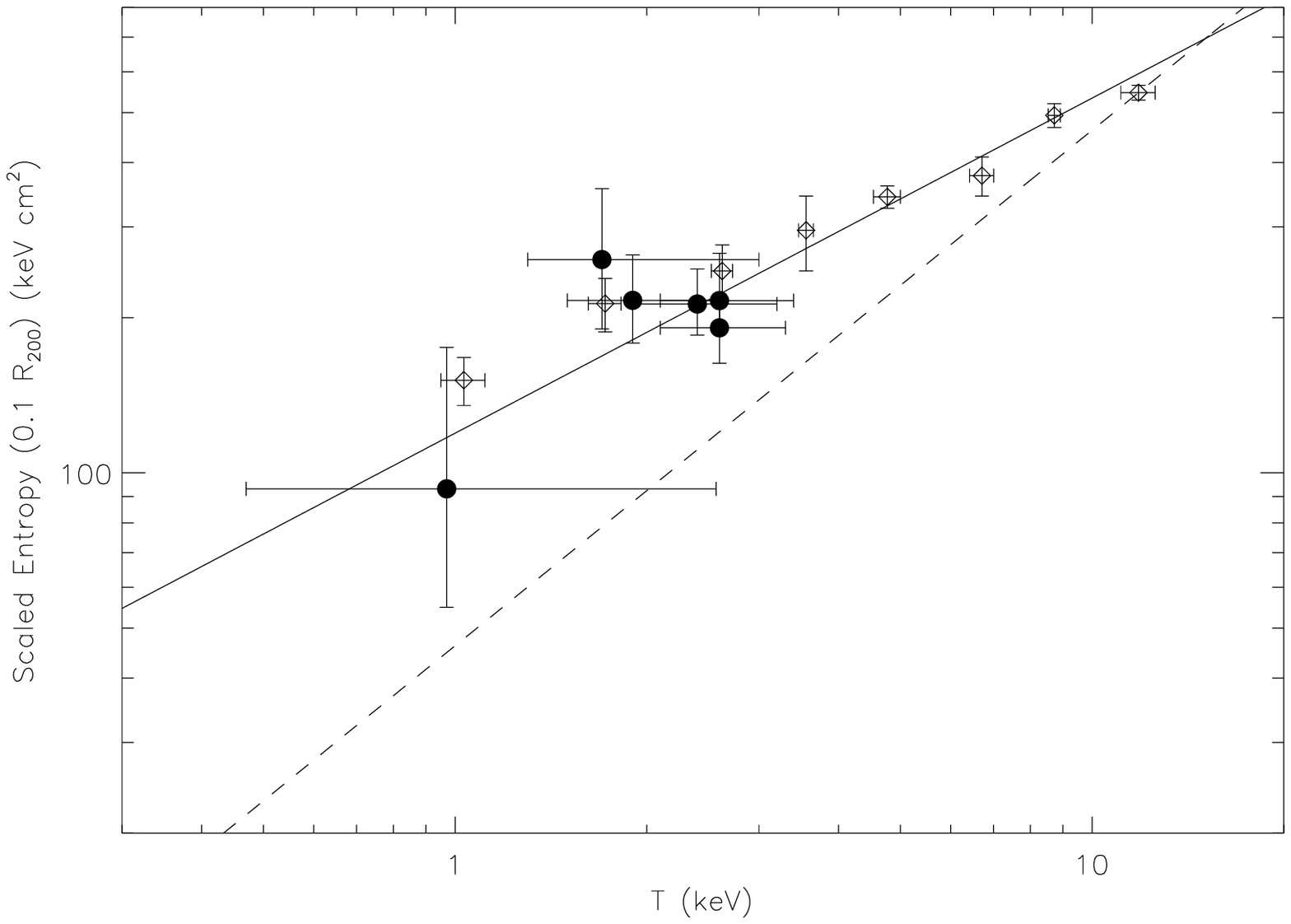}
\figcaption{ Entropy at a radius of $0.1r_{200}$ versus temperature. We multiply the entropies of our groups by $E_z^{4/3}$ to account for the variation of the mean density within a given overdensity radius with redshift.  These points are plotted with filled circles.  Errors are $1\sigma$ in entropy and 90\% in temperature.  Open diamonds show the entropies of the low-redshift cluster and group sample of Ponman et al. (2003) grouped to give a minimum of eight clusters per bin.  The solid line shows the best-fit to the Ponman et al. (2003) sample, and the dashed line shows the predicted self-similar scaling of entropy with temperature normalized to the eight hottest clusters. }
\end{figure}

\begin{figure}
\centering
\epsscale{1.0}
\plotone{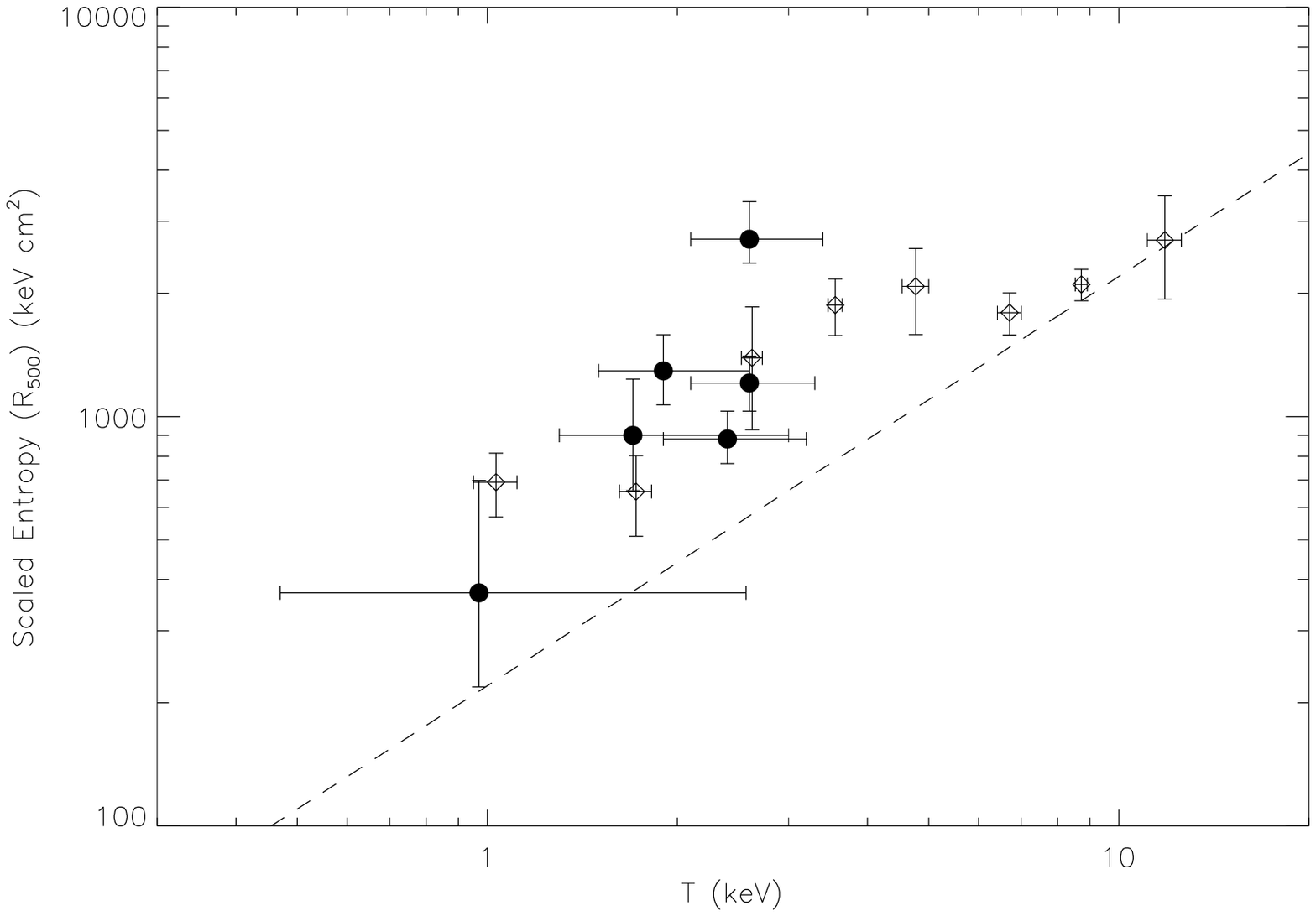}
\figcaption{ Same as Figure 12 for entropy within a radius of $r_{500}$. }
\end{figure}

\section{ AGN IN GROUPS }

Several studies with Chandra have detected an overdensity of X-ray
sources toward both individual clusters (Cappi et al. 2001; Sun \&
Murray 2002; Molnar et al. 2002) and cumulatively in large cluster
samples (Cappelluti et al. 2005; Ruderman \& Ebeling 2005) relative to
the field.  Through optical spectroscopy of the X-ray sources in eight
clusters, Martini et al. (2006) securely identified 40 luminous, X-ray
sources ($L_X > 8 \times 10^{40}$ ergs s$^{-1}$) with cluster member
galaxies.  At these luminosities most of these sources, particularly
those with $L_X > 10^{42}$ ergs s$^{-1}$, are AGN, although only four
of these galaxies show AGN signatures in the optical spectra.  These
observations reveal a population of AGN associated with clusters that
have gone previously unidentified in optical observations.

Very little is known about AGN populations in low-mass clusters and
groups.  Only a few of the X-ray sources in our group fields
correspond to galaxies for which we have spectroscopy.  One source in
RXJ1648+60, XMMU J164838.1+601934, matches a group member galaxy at
$z=0.3756$.  This source has a hard band (2-10 keV) luminosity of $1.3
\times 10^{43}$ ergs s$^{-1}$ and a broad band (0.3-8 keV) luminosity
of $2.0 \times 10^{43}$ ergs s$^{-1}$, securely identifying it as an
AGN.  The approximately 200 counts in the XMM observation allow for a
rough spectral fit, which gives a power law index of
$1.7^{+0.8}_{-0.6}$ and which is consistent with galactic absorption.
Through comparison to our optical imaging, we find a number of other
candidate group AGN.  Within a radius of 1 Mpc there are six X-ray
sources matching galaxies bright enough to host AGN.  Assuming these
sources are at the group redshifts and have typical AGN spectra with a
power law index of 1.7 and galactic absorption, they all have X-ray
luminosities of $10^{42}$ ergs s$^{-1}$ or greater.

We also investigate whether or not there is an overdensity of X-ray
sources in our group fields compared to the observed Log$N$-Log$S$
from studies of the cosmic X-ray background (Moretti et al. 2003;
Rosati et al. 2002; Tozzi et al. 2001; Brandt et al. 2001; Mushotsky
2000).  For this calculation we include only X-ray sources within 1
Mpc of the group center and with fluxes greater than $9 \times
10^{-15}$ ergs cm$^{-2}$ s$^{-1}$ in the hard band (2-10 keV) and $5
\times 10^{-15}$ ergs cm$^{-2}$ s$^{-1}$ in the soft band (0.5-2 keV).
These flux limits ensure that the sources could be detected in all of
our group exposures.  In our six group fields, we detect 25 sources
which meet these requirements.  In comparison, fits to the
Log$N$-Log$S$ predict 13-16 sources in the soft band and 19-26 sources
in the hard band, depending on the study (Moretti et al. 2003; Rosati
et al. 2002; Tozzi et al. 2001; Brandt et al. 2001; Mushotsky 2000).
We therefore find approximately a 2$\sigma$ excess in the number of
soft band sources and no significant excess in the number of hard band
sources.  However, we are only probing the bright end of the
Log$N$-Log$S$.

In summary, we identify at least one X-ray luminous AGN associated
with one of our groups, and find indications for a population of group
AGN similar to those seen in clusters.  Deeper X-ray observations
combined with spectroscopy of the optical counterparts of the X-ray
sources would confirm the existence of a significant fraction of AGN
in these groups.

\section{ SUMMARY }

A lot of work has been done on the largest collapsed objects in the
universe, massive clusters of galaxies, with studies now extending
beyond a redshift of one.  However, little is known about their more
common low-mass counterparts, poor clusters and groups of galaxies,
beyond the present epoch.  In addition to forming the building blocks
of larger structures, groups contain most of the galaxies in the
universe and are likely the sites of significant galaxy evolution
(e.g. Tully 1987).  As part of a multiwavelength study of moderate
redshift groups, we obtained \textit{XMM-Newton} observations of six
groups with redshifts between 0.2 and 0.6.

We find generally good agreement between the X-ray properties of our
groups and those at lower redshifts.  Similar to low-redshift groups,
the X-ray emission in several of our groups is centered on a dominant
early-type galaxy, and in the case of RXJ0329+02, the position angle
of the X-ray isophotes aligns with the central galaxy.  However, two
of our groups are not centered on a BGG, indicating they are probably
less dynamically evolved (Paper I).  RXJ1334+37 has a BGG, but the X-ray
emission peaks to the south-east of this galaxy.  RXJ1648+60 contains
a string of bright galaxies which are traced by the X-ray emission
rather than one dominant galaxy.  The X-ray morphology of RXJ1334+37
is also less symmetric than the other groups in our sample.  The
contamination by background flares makes it difficult to determine the
structure of RXJ1648+60 from the current observation, but it may also
be elongated.

Our groups have temperatures around 2 keV placing them in the massive
group or poor cluster regime of galaxy associations.  Their X-ray
properties are in agreement with the scaling relations between
luminosity, temperature, and velocity dispersion defined by
low-redshift groups and clusters as well as with the shallower
observations of intermediate-redshift groups in the XMM-LSS (Osmond \&
Ponman 2004; Horner 2001; Willis et al. 2005).  Our groups appear to
be in slightly better agreement with the cluster scaling relations;
however, they all agree within the errors.  In particular, there is a large
scatter in the velocity dispersions of our groups with a couple of groups
having velocity dispersions that are too low for their temperatures and 
luminosities.  
The spectrum of RXJ1648+60 only has a S/N of 3.6, and the 
associated error in the temperature could make it consistent with the 
$\sigma_v-T_X$ relation but not the the $L_X-\sigma_v$ relation (Paper I).
This discrepancy could be an artifact
of the X-ray selection or a small number of velocity measurements, but
their velocity dispersions lie close to the theoretical lower limit
for collapsed systems (Mamon 1994), which is difficult to explain in
systems which clearly show extended X-ray emission.  Similar low
velocity dispersions are observed in a few low-redshift groups (Osmond
\& Ponman 2004; Helsdon et al. 2005).  The proposed explanations
include a lowering of the velocity dispersion through dynamical
friction, tidal heating, or orientation effects.  Future deeper
spectroscopy will help us to understand these systems.

Our observations indicate that intermediate-redshift groups contain
excess entropy over the expected self-similar scaling with
temperature.  We find this excess at both small ($0.1r_{200}$) and
large ($r_{500}$) radii, similar to a study of 66 clusters and groups
at low-redshift (Ponman et al. 2003).  The detection of excess entropy
out to large radii is significant, because models like preheating
generally predict that the excess entropy should be restricted to the
central regions.  Finally, we detect at least one X-ray luminous AGN
associated with a group member galaxy.  This source is associated with
RXJ1648+60 and has a broad band (0.3-8 keV) luminosity of $2.0 \times
10^{43}$ ergs s$^{-1}$.  We also find several other luminous X-ray
point sources in these groups which match galaxies in the optical
imaging and a 2$\sigma$ excess in the number of soft band (0.5-2 keV)
sources over the field.  While not conclusive, these findings may
point to a population of group AGN similar to those seen in clusters
(Martini et al. 2006).

\acknowledgements

We would like to thank D. Horner, A. Sanderson, and J. Osmond for
sharing information about their results and T. Ponman for providing us with the GEMS fits.  
JSM acknowledges support from NASA grants NNG04GC846 and NNG04GG536.

\begin{deluxetable}{lccccc}
\tablecaption{ Group Sample }
\tablewidth{0pt}
\tablecolumns{6}
\tablehead{
\colhead{Group} & \colhead{R.A. (J2000)} & \colhead{Decl. (J2000)} & \colhead{Exposure (ks)} &
\colhead{Redshift} & \colhead{$\sigma$}\\
\colhead{} & \colhead{} & \colhead{} & \colhead{MOS1, MOS2, PN} &
\colhead{} & \colhead{(km s$^{-1}$)}
}
\startdata
 RXJ0329.0+0256 &03:29:02.82 &+02:56:25.2 &46, 46, 39 &0.412 &258$^{+95}_{-46}$ \\
 RXJ0720.8+7109 &07:20:54.04 &+71:08:57.9 &8.5, 8.4, 5.7 &0.231 &620$^{+93}_{-53}$ \\
 RXJ1205.9+4429 &12:05:51.44 &+44:29:11.0 &24, 24, 17 &0.593 &530$^{+407}_{-406}$ \\
 RXJ1256.0+2556 &12:56:02.34 &+25:56:37.1 &23, 23, 18 &0.232 &656$^{+93}_{-57}$ \\
 RXJ1334.0+3750 &13:34:58.95 &+37:50:15.7 &104, 76, 60  &0.384 &121$^{+58}_{-45}$ \\
 RXJ1648.7+6019 &16:48:43.63 &+60:19:21.5 &20, 23, 7.2 &0.376  &130$^{+46}_{-48}$\\
\enddata
\tablecomments{ Column 4 lists the exposure times after the removal of background flares.  For RXJ1648+60 and RXJ1334+37, the total exposure times from both observations are listed. Errors in the velocity dispersion are $1\sigma$. }
\end{deluxetable}

\begin{deluxetable}{lcccc}
\tablecaption{ Spatial Properties }
\tablewidth{0pt}
\tablecolumns{5}
\tablehead{
\colhead{Group} & \colhead{$r_{core}$ (arcsec, kpc)} & \colhead{$\beta$} & \colhead{Ellipticity} & \colhead{P.A.}
}
\startdata
 RXJ0329.0+0256 &$63^{+53}_{-17}$, $348^{+292}_{-94}$ &$1.2^{+1.5}_{-0.3}$ &$0.37^{+0.07}_{-0.07}$ &$-59\pm6$ \\
 RXJ0720.8+7109 &$45^{+26}_{-11}$, $167^{+96}_{-41}$ &$0.91^{+0.62}_{-0.18}$ &$<0.012$ &- \\
 RXJ1205.9+4429 &$18^{+5}_{-4}$, $121^{+34}_{-27}$ &$0.67^{+0.11}_{-0.09}$ &$<0.14$ &- \\
 RXJ1256.0+2556 &$9^{+2}_{-3}$, $33^{+7}_{-11}$ &$0.40^{+0.03}_{-0.02}$ & $<0.0086$ &- \\
 RXJ1334.0+3750 &$26^{+12}_{-8}$, $137^{+64}_{-42}$ &$0.54^{+0.11}_{-0.06}$ &$0.51^{+0.06}_{-0.08}$ &$-41^{+6}_{-5}$ \\
 RXJ1648.7+6019 &$<62$, $<324$ &$>0.50$ &-  &- \\
\enddata
\tablecomments{ Errors and limits are $1\sigma$.  Due to the lack of counts for RXJ1648+60, we did not fit for ellipticity, and we fixed the background using the observed local background. For RXJ1334+37 the background was also fixed at a local background, and the center was fixed at the X-ray peak/centroid. The last column gives the position angle of the semi-major axis, measured counterclockwise from north, for the two elliptical groups. }
\end{deluxetable}

\begin{deluxetable}{lcccccccc}
\tablecaption{ Spectral Properties }
\tablewidth{0pt}
\tablecolumns{9}
\tablehead{
\colhead{Group} & \colhead{Redshift} & \colhead{$r_{ext}$} & \colhead{S/N} & \colhead{kT} &
\colhead{$Z$} & \colhead{$L_X$} & \colhead{$r_{500}$} & \colhead{$L_X$($r_{500}$)}\\
\colhead{} & \colhead{} & \colhead{} & \colhead{} & \colhead{(keV)} & \colhead{($Z_{\odot}$)} & \colhead{($10^{43}$ ergs s$^{-1}$)} & \colhead{(kpc)} & \colhead{($10^{43}$ ergs s$^{-1}$)}
}
\startdata
 RXJ0329+02&0.412&96''&13&$1.9^{+0.7}_{-0.4}$&$0.20^{+0.56}_{-0.20}$&$6.0^{+2.3}_{-1.3}$&691&$6.3^{+2.4}_{-1.4}$ \\
 RXJ0720+71&0.231&104''&13&$2.6^{+0.8}_{-0.5}$&$0.20^{+0.54}_{-0.20}$&$5.3^{+1.8}_{-1.0}$&850&$5.8^{+2.0}_{-1.1}$ \\
 RXJ1205+44&0.593&80''&14&$2.6^{+0.7}_{-0.5}$&$0.72^{+1.28}_{-0.45}$&$15^{+7}_{-3}$&599&$15^{+8}_{-3}$ \\
 RXJ1256+25&0.232&156''&16&$2.4^{+0.8}_{-0.5}$&$0.02^{+0.22}_{-0.02}$&$3.8^{+0.9}_{-0.4}$&550&$3.7^{+0.8}_{-0.4}$ \\
 RXJ1334+37&0.384&48''&10&$1.7^{+1.3}_{-0.4}$&$0.50^{+6.95}_{-0.42}$&$0.95^{+0.30}_{-0.29}$&479&$1.5^{+0.5}_{-0.4}$ \\
 RXJ1648+60&0.376&36''&3.6&$0.97^{+1.59}_{-0.54}$&(0.3)&$1.7^{+1.8}_{-1.0}$&413&$2.8^{+3.0}_{-1.6}$ \\
\enddata
\tablecomments{ The errors listed are 90\% confidence limits.  Column 4 lists the signal-to-noise of the group spectra calculated by summing the counts in the three detectors.  For RXJ1648+60, the metallicity was fixed at 0.3 solar.  The listed luminosities are unabsorbed, bolometric luminosities.  They are corrected for the area lost from point source removal, and the luminosity errors take into account both the uncertainties in temperature and metallicity. }
\end{deluxetable}

\end{document}